\documentclass[traditabstract]{aa}

\usepackage{amsmath}
\usepackage{amssymb}
\usepackage{upgreek}
\usepackage[varg]{txfonts}
\usepackage{epsfig,graphicx}
\usepackage{color}
\usepackage[colorlinks=true, linkcolor=blue, citecolor=blue, urlcolor=blue]{hyperref}
\usepackage{natbib}
\bibpunct{(}{)}{;}{a}{}{,} 
\newcommand{\txd}{{\text{d}}}
\newcommand{\calE}{{\cal{E}}}
\renewcommand{\Re}{R_{\text{e}}}
\newcommand{\Ie}{I_{\text{e}}}

\defcitealias{1991A&A...249...99C}{Paper~I}
\defcitealias{1997A&A...321..724C}{Paper~II}

\begin{document}

\title{Stellar systems following the $R^{1/m}$ luminosity law}
\subtitle{III. Photometric, intrinsic, and dynamical properties for all S\'ersic indices}

\author{Maarten Baes\inst{\ref{UGent}} \and Luca Ciotti\inst{\ref{Bologna}}}

\institute{
Sterrenkundig Observatorium, Universiteit Gent, Krijgslaan 281 S9, 9000 Gent, Belgium
\label{UGent}
\and
Dipartimento di Fisica e Astronomia, Universit\`{a} di Bologna, Via Piero Gobetti 93/2, Bologna, Italy
\label{Bologna}
}

\date{Received 20 February 2019 / Accepted 24 May 2019}

\abstract{The S\'ersic or $R^{1/m}$ model has become the de facto standard model to describe the surface brightness profiles of early-type galaxies and the bulges of spiral galaxies. The photometric, intrinsic, and dynamical properties of this model have been investigated, but mainly for fairly large S\'ersic indices $m$. For small values of $m$, appropriate for low-mass and dwarf ellipticals, a detailed investigation of these properties is still lacking. In this study, we used a combination of numerical and analytical techniques to investigate the S\'ersic model over the entire range of S\'ersic parameters, focussing on the small $m$ regime, where a number of interesting and surprising properties are found. For all values $m<1$, the model is characterised by a finite central luminosity density, and for $m<\tfrac12$, even a central depression in the luminosity density profile. This behaviour translates to the dynamical properties: we show that all S\'ersic models with $m \geqslant\tfrac12$ can be supported by an isotropic velocity dispersion tensor, and that these isotropic models are stable to both radial and non-radial perturbations. The models with $m < \tfrac12$, on the other hand, cannot be supported by an isotropic velocity dispersion tensor.}

\keywords{galaxies: photometry -- galaxies: structure -- galaxies: kinematics and dynamics -- methods: analytical}

\maketitle

\section{Introduction}

The construction of simple parametric models that can still provide a realistic representation of galaxies is important for various reasons. Such models enable the characterisation of galaxies by a number of relevant physical parameters, which can be used for galaxy classification or to study scaling relation studies. Realistic yet simple parametric models are also crucial as starting points for detailed theoretical studies or numerical simulations, including N-body simulations, hydrodynamics simulations or radiative transfer simulations.

Most analytical models for galaxies start from a parameterised model for the density distribution. Many of the most commonly used models belong to this category, such as the Plummer sphere \citep{1911MNRAS..71..460P, 1987MNRAS.224...13D} , the isochrone sphere \citep{1959AnAp...22..126H}, the Hernquist model \citep{1990ApJ...356..359H, 2002A&A...393..485B}, the Jaffe model \citep{1983MNRAS.202..995J}, or the families of $\gamma$ models \citep{1993MNRAS.265..250D, 1994AJ....107..634T} or Einasto models \citep{1965TrAlm...5...87E, 2005MNRAS.358.1325C}. These models have an advantage allowing many important intrinsic and dynamical quantities, including the density, gravitational potential, velocity dispersion profiles, and the phase-space distribution function, to be calculated analytically. On the other hand, the surface brightness distribution of these models is not always a reliable representation for the observed surface brightness distribution of real galaxies. 

To avoid this problem, one can consider a parameterised function for the surface brightness distribution, and construct 3D models by de-projecting this distribution. A prime example of a parameterised surface brightness distribution is the formula proposed by \citet{1948AnAp...11..247D},
\begin{equation}
I(R) = \Ie \exp\left\{-7.669\left[\left(\frac{R}{\Re}\right)^{1/4}-1\right]\right\},
\end{equation}
generally known as the $R^{1/4}$ or the de Vaucouleurs model. It is characterised by only two parameters: the effective radius $\Re$ and the surface brightness $\Ie$ at the effective radius. The de Vaucouleurs model was introduced to describe the surface brightness profiles of elliptical galaxies, and turned out to be a remarkably good model for many observed galaxies. One famous example is the massive elliptical galaxy NGC\,3379, for which the model proved an excellent fit over more than 10 magnitudes in surface brightness \citep{1979ApJS...40..699D}. A drawback of this model is that many of the intrinsic properties, such as the density and gravitational potential, cannot be expressed analytically. These properties were investigated through numerical means in detail \citep{1960BOTT....2t...3P, 1976AJ.....81..807Y, 1982MNRAS.200..951B}.

It soon became clear, however, that not all early-type galaxies are adequately fitted by a de Vaucouleurs model. A generalisation of the de Vaucouleurs model was presented by \citet{1968adga.book.....S}, and is known as the $R^{1/m}$ or S\'ersic model. It is characterised by the surface brightness profile
\begin{equation}
I(R) = \Ie \exp\left\{-b\left[\left(\frac{R}{\Re}\right)^{1/m}-1\right]\right\},
\label{SersicI}
\end{equation}
with $m$ the so-called S\'ersic index, and $b\equiv b(m)$ a dimensionless parameter that depends only on the value of $m$. Over the past decades, the S\'ersic model has become the preferred model to describe the surface brightness profiles of early-type galaxies, as well as the bulges of spiral galaxies \citep[e.g.,][]{1988MNRAS.232..239D, 1993MNRAS.265.1013C, 2001A&A...368...16M, 2003AJ....125.2936G, 2006MNRAS.371....2A, 2009MNRAS.393.1531G, 2012MNRAS.427.1666B, 2012ApJS..203...24V, 2015ApJS..219....4S, 2016MNRAS.462.1470L}.

Given the popularity of this model, the photometric, intrinsic, and dynamical properties of the S\'ersic model have been examined in quite some detail (\citealt{1991A&A...249...99C}, hereafter \citetalias{1991A&A...249...99C}; \citealt{1997A&A...321..724C}, hereafter \citetalias{1997A&A...321..724C}; \citealt{1999A&A...352..447C, 2002A&A...383..384M, 2002MNRAS.333..510T, 2004A&A...415..839C, 2005PASA...22..118G, 2007JCAP...07..006E, 2011A&A...525A.136B, 2011A&A...534A..69B}). In \citetalias{1991A&A...249...99C} we investigated the most important intrinsic and dynamical properties of the family of S\'ersic models. A fundamental result from this study was that the S\'ersic models can be supported by an isotropic velocity dispersion tensor, and these isotropic models are stable to both radial and non-radial perturbations.

One limitation of this study, and of most other theoretical studies devoted to the family of S\'ersic models, is that it is limited to models with relatively large S\'ersic indices, $2\leqslant m\leqslant10$. Models with smaller S\'ersic indices were not considered, as the S\'ersic model was mainly applied to massive early-type galaxies at that time, and this range of parameters is fully appropriate for that regime. For less massive galaxies, however, smaller values of $m$ are more appropriate. Indeed, the S\'ersic index is found to correlate systematically with various galaxy parameters, including central surface brightness, luminosity, and effective radius, with larger indices ($m\gtrsim2$) typical for massive elliptical galaxies, and smaller indices ($m\lesssim2$) found in lower-mass and dwarf ellipticals \citep{1993MNRAS.265.1013C, 1994MNRAS.271..523D, 1994MNRAS.268L..11Y, 1998A&A...333...17B, 2003AJ....125.2936G, 2004ApJ...602..664G}. In fact, several studies found models with S\'ersic indices down to values $m\approx0.4$ \citep[e.g.,][]{1994MNRAS.268L..11Y, 1998A&A...333...17B, 2003ApJ...582..689M}. 

In this paper, we extend our previous studies in this series \citepalias{1991A&A...249...99C, 1997A&A...321..724C} to S\'ersic models with small S\'ersic indices, focussing on the regime $m\leqslant1$, and even covering the limit $m\rightarrow0$. We study the systematic behaviour of the most important intrinsic and dynamical properties, using a combination of numerical and analytical techniques. In Sect.~{\ref{Photometric.sec}} we discuss the photometric and intrinsic properties of the family of models, with particular attention for the S\'ersic scaling parameter and the spatial luminosity density for models with $m\leqslant1$. In Sect.~{\ref{Dynamical.sec}} we focus on the most important dynamical properties, including velocity dispersion profiles, the distribution function, and the differential energy distribution. We show throughout this paper that the results for $m>1$ cannot be generalised to smaller S\'ersic indices, and instead find interesting behaviour. We summarise our findings in Sect.~{\ref{Summary.sec}}.

\section{Photometric properties}
\label{Photometric.sec}

\subsection{Surface brigthness profile}
\label{surf.sec}

\begin{figure*}
\centering
\includegraphics[width=\textwidth]{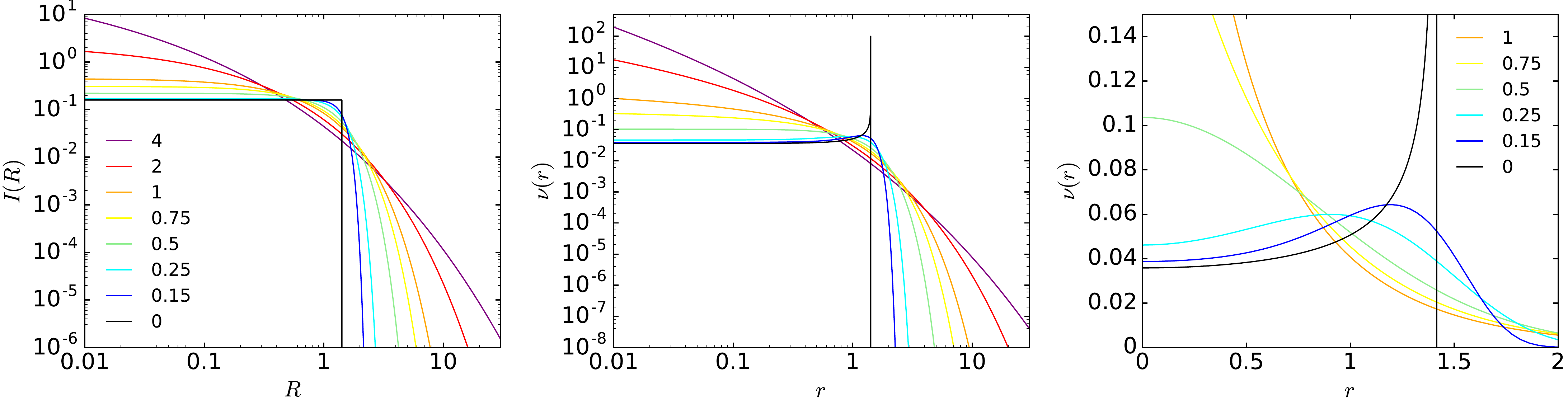}
\caption{Left: Surface brightness profile $I(R)$ for S\'ersic models with different values of $m$, focussing on models with small $m$. In this and all of the following figures, we work in dimensionless units, or equivalently, we consider a model with $\Re=L=M=1$. Middle: Luminosity density profile $\nu(r)$ for the same set of models. Right: Luminosity profiles of models with $m\leqslant1$, but now shown in linear scale instead of log-log scale. All model with $m<\tfrac12$ are characterised by a central depression.}  
\label{SmallSersic-I-nu.fig}
\end{figure*}

The family of S\'ersic models is defined by the surface brightness profile (\ref{SersicI}). When we replace the effective surface brightness by the total luminosity $L$ as a free parameter, we find
\begin{equation}
I(R) = \frac{b^{2m}}{2\pi\,m\,\Gamma(2m)}\,\frac{L}{\Re^2}\exp\left[-b\left(\frac{R}{\Re}\right)^{1/m}\right].
\label{sersic}
\end{equation}
It is a three-parameter family with the total luminosity $L$ and the effective radius $\Re$ as the length and luminosity scales, and the S\'ersic index $m$ a parameter that controls the shape of the surface brightness profile. The parameter $b = b(m)$ in this formula is not a free parameter in the model, but a dimensionless scaling parameter that  depends on $m$, and is such that $\Re$ corresponds to the isophote that contains half of the emitted luminosity (we will discuss this parameter in more detail in Sect.~{\ref{bm.sec}}). In the remainder of this paper, we will generally keep $L$ and $\Re$ fixed, and study the behaviour of the S\'ersic model as a function of the S\'ersic index $m$.

All S\'ersic models are characterised by a finite central surface brightness. For given values of $L$ and $\Re$, the central surface brightness is an increasing value of the S\'ersic index $m$. As we will see in the next subsection, the parameter $b(m)$ tends to zero for $m\rightarrow0$ in a way that the central intensity converges to 
\begin{equation}
I_0 = \frac{L}{2\pi\,\Re^2}.
\end{equation}
The fact that the total luminosity and the central surface brightness are finite even in the limit $m\rightarrow0$ suggests that this limiting model is worth looking at in more detail. The left panel of Fig.~{\ref{SmallSersic-I-nu.fig}} shows the intensity of the S\'ersic profile for different values of $m$. For increasingly smaller values of $m$, the S\'ersic intensity profile tends to approach a constant value at small radii, and then suddenly and quickly drops to zero beyond a certain radius. By using equation~(\ref{b2mseries}) it follows that, in the limit $m\rightarrow0$, the S\'ersic model reduces to a simple model characterised by a finite extent and a constant surface brightness,
\begin{equation}
I(R) = \begin{cases}
\;I_0 &R<\!\sqrt{2}\,\Re, \\[0.1em] \;0&R>\!\sqrt{2}\,\Re.
\end{cases}
\label{Sigma0}
\end{equation}

\subsection{Cumulative surface brightness and scaling parameter}
\label{bm.sec}

For the S\'ersic model, the cumulative surface brightness $S(R)$, that is the integrated flux emitted within the isophote at radius $R$,
\begin{equation}
S(R) = 2\pi \int_0^R I(R')\,R'\,{\text{d}}R',
\end{equation}
can conveniently be expressed in terms of the incomplete gamma function
(\citetalias{1991A&A...249...99C}; \citealt{2005PASA...22..118G}),
\begin{equation}
\frac{S(R)}{L} =
\frac{\gamma\left[2m,b\left(\dfrac{R}{\Re}\right)^{1/m}\right]}{\Gamma(2m)}.
\label{LR}
\end{equation}
In the limit $m\rightarrow0$ this reduces to 
\begin{equation}
\frac{S(R)}{L} = \begin{cases}
\;\dfrac{R^2}{2\Re^2} &R<\!\sqrt{2}\,\Re, \\[0.5em] \;1&R>\!\sqrt{2}\,\Re. \end{cases}
\end{equation}
As indicated in the previous section, the dimensionless scaling parameter $b$ is not a free parameter in the model, but a dimensionless number that depends on the value of $m$. Combining equation~(\ref{LR}) with the definition of $\Re$ as the isophote that contains half of the emitted luminosity, it is easy to see that the value of $b$ can be found by solving the non-algebraic equation
\begin{equation}
\gamma(2m,b) = \frac{\Gamma(2m)}{2}.
\label{solvebm}
\end{equation}
Various interpolation formulae for $b$ at large $m$ have been presented in the literature (\citealt{1989woga.conf..208C}; \citetalias{1991A&A...249...99C}; \citealt{1997A&A...321..111P}). \citet{1999A&A...352..447C} provide a full asymptotic expansion for $b$ for $m\rightarrow\infty$, and demonstrate that the truncated series with terms up to order $m^{-4}$ forms an excellent approximation, even for values of $m$ as small as one. 

\begin{figure*}
\centering
\includegraphics[width=\textwidth]{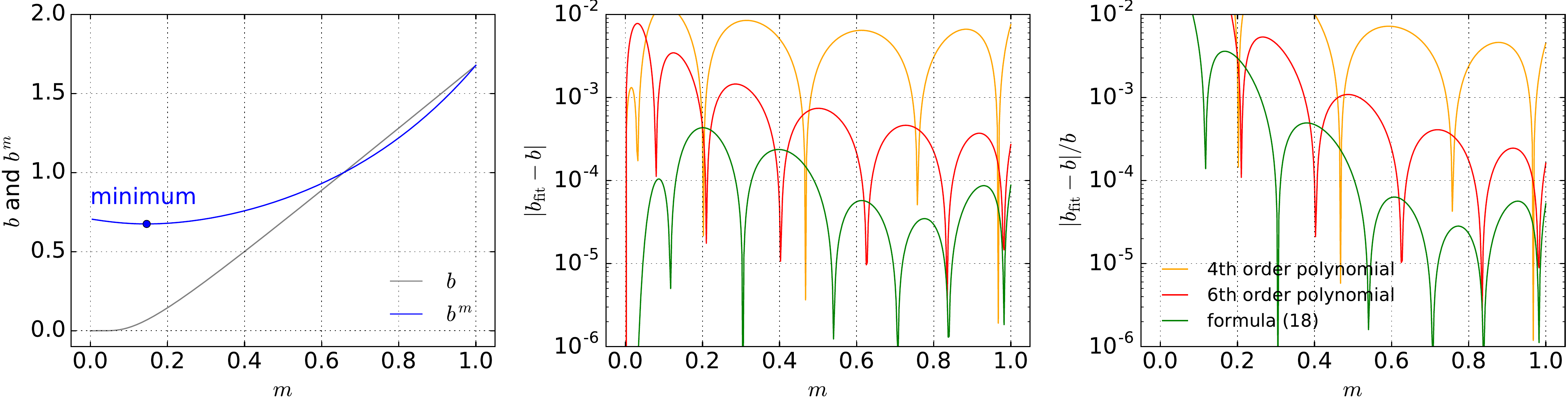}
\caption{Left: S\'ersic scaling parameter $b$ and $b^m$ as a function of $m$ in the range $0\leqslant m\leqslant1$. The minimum value for $b^m$ is obtained at $m\approx0.146$. Centre and right: Absolute accuracy and relative accuracy of different fitting formulae to $b$ in the range $0\leqslant m\leqslant1$.}
\label{ScalingParameterFits.fig}
\end{figure*}

For the small $m\leqslant1$ values we focus on in this paper, however, this expansion is not applicable . This was also noted by \citet{2003ApJ...582..689M}, who propose a separate interpolation formula for small $m$. In Table~{\ref{bm.tab}} we provide accurate values for $b$ for values of $m$ between 0 and 1, obtained by solving equation (\ref{solvebm}) numerically. The variation of $b$ as a function of $m$ in this range is shown in the left panel for Fig.~{\ref{ScalingParameterFits.fig}}. Somewhat surprisingly, it turned out impossible to find a good polynomial interpolation formula for $b(m)$. The orange and red curves in central and right panels of Fig.~{\ref{ScalingParameterFits.fig}} show the absolute and relative accuracy for the fourth and sixth order polynomial fits; no polynomial fit can reach absolute accuracy better than $10^{-3}$ over the interval $0\leqslant m\leqslant 1$, or a relative accuracy better than $10^{-3}$ over the interval $0.2\leqslant m\leqslant 1$ (relative errors are less meaningful for $m\rightarrow0$ as $b(m)$ tends to zero in this limit).

In order to find a suitable fitting formula, one can argue that $b^m$ rather than $b$ is the fundamental scaling parameter in the S\'ersic model. In most of the expressions for integrated quantities, the scaling parameter appears only in the combination $b^m$. From a mathematical point of view, the S\'ersic surface brightness formula (\ref{sersic}) can be represented in a more `natural' way as
\begin{equation}
I(R) = \frac{1}{2\pi\,m\,\Gamma(2m)}\,\frac{L}{h^2}\exp\left[-\left(\frac{R}{h}\right)^{1/m}\right],
\end{equation}
with the scale length $h$ instead of the effective radius as the obvious length scale \citep{1979MNRAS.187..357E, 1988MNRAS.232..239D}. The parameter $b^m$ is then nothing but the ratio $\Re/h$. 

The third column in Table~{\ref{bm.tab}} lists $b^m$ for S\'ersic indices $m$ down to zero. Since $b$ converges to zero for $m\rightarrow0$, the combination $b^m$ seems ill-defined. However, using the standard technique of asymptotic expansion, we prove in Appendix~{\ref{bmzero.sec}} that the leading term of the asymptotic expansion of $b(m)$ to $m\to0$ can be written in closed form as
\begin{equation}
b^m \sim \sqrt{\,m\,\Gamma(2m)}.
\label{bmzeroapp}
\end{equation}
In particular, this implies that $b^m$ converges to the finite value 
\begin{equation}
\lim_{m\rightarrow0} b^m = \frac{1}{\sqrt2}.
\end{equation}
This result immediately proves the equations in Sect.~{\ref{surf.sec}} in the limit $m\rightarrow0$.

The expression (\ref{bmzeroapp}) only provides an accurate approximation for very small values of $m$, and it is not very useful as a practical approximation for general $m\leqslant1$. However, it turns out that, contrary to $b$ itself, $b^m$ is very well fitted by a polynomial approximation over the interval $0\leqslant m\leqslant 1$. Indeed, the formula
\begin{gather}
b^m \approx \frac{1}{\sqrt{2}} + \sum_{k=1}^4 a_k\,m^k, \\
a_1 = -0.45807, \nonumber \\
a_2 = 1.83247, \nonumber \\
a_3 = -1.2556, \nonumber \\
a_4 = 0.85239, \nonumber
\end{gather}
provides a fit with a mean absolute accuracy of $1.6\times10^{-4}$ and a mean relative accuracy of $2.2\times10^{-4}$ over the interval $0\leqslant m\leqslant 1$. The corresponding fitting formula for $b$, 
\begin{equation}
b \approx \left(\frac{1}{\sqrt{2}} + \sum_{k=1}^4 a_k\,m^k\right)^{1/m},
\end{equation}
is shown as the green curve in Fig.~{\ref{ScalingParameterFits.fig}}. It is clearly a superior approximation to $b$ than the polynomial fits. Over the entire interval $0\leqslant m\leqslant1$, it is characterised by a mean absolute accuracy of $1.1\times10^{-4}$.

Curiously, the quantity $b^m$ does not monotonically increase as a function of $m$: the minimum value for $b^m$ is found for $m\approx0.146$.

\subsection{Luminosity density}
\label{density.sec}

Starting from a surface brightness profile $I(R)$, one can calculate the deprojected emissivity or luminosity density $\nu(r)$  through an inverse Abel transform. The general expression, valid for intensity profiles with an infinite as well as a finite extent, is \citep[e.g.,][]{2000stdy.book.....C},
\begin{equation}
\nu(r) = -\frac{1}{\pi}\int_r^{r_{\text{t}}} \frac{{\text{d}}I(R)}{{\text{d}}R}\,\frac{{\text{d}}R}{\sqrt{R^2-r^2}}
+ {I(r_{\text{t}})\over \pi\sqrt{r_{\text{t}}^2 -r^2}}
\label{deprojection-gen}
\end{equation}
with $r_{\text{t}}$ the truncation radius. For models with an infinite extent, $r_{\text{t}}\to\infty$ and the formula simplifies to \citep{2008gady.book.....B}
\begin{equation}
\nu(r) = -\frac{1}{\pi}\int_r^\infty \frac{{\text{d}}I(R)}{{\text{d}}R}\,\frac{{\text{d}}R}{\sqrt{R^2-r^2}}.
\label{deprojection}
\end{equation}
It is well-known that there is no closed expression for the density corresponding the S\'ersic model in terms of elementary functions or standard special functions. \citet{2011A&A...534A..69B} used Mellin transforms to derive a general expression for the density and some related properties of the S\'ersic model in terms of the Fox $H$ function\footnote{The Fox $H$ function is a special function that is a powerful tool for analytical work. It is gradually becoming more adopted in applied sciences, including astrophysics \citep[e.g.,][]{2006Ap&SS.305..289S, 2007BASI...35..681H, 2009ApJ...690.1280V, 2012A&A...546A..32R, 2012A&A...540A..70R, 2012RMxAA..48..209Z}. The definition and some interesting properties of the Fox $H$ function can be found in the appendices of \citet{2009ApJ...690.1280V} and \citet{2011A&A...534A..69B}; for a full overview of the many useful properties of these extremely flexible special functions, we refer to the comprehensive work by \citet{2009hfta.book.....M}.}. The general expression reads
\begin{equation}
\nu(r) =
\frac{b^{3m}}{\pi^{3/2}\,\Gamma(2m)}\,\frac{L}{\Re^3}\,
u^{-1}\,H^{2,0}_{1,2} \left[ \left.\begin{matrix} (0,1) \\ (0,2m), (\tfrac12,1) \end{matrix} \,\right| u^2 \right],
\label{rho-FoxH}
\end{equation}
where we have introduced the reduced radial coordinate
\begin{equation}
u = \dfrac{b^m r}{\Re} = \frac{r}{h}.
\end{equation}
This expression is maybe not directly useful for numerical calculations, because the general Fox $H$ function is not (yet) implemented in many numerical libraries or software packages. Implementations have been presented though for various programming languages, including C and MATLAB \citep{2018arXiv180408101C}, Mathematica \citep{2012arXiv1202.2576S}, and Python \citep{Alhennawi2016}. For rational $m$, the Fox $H$ function in expression (\ref{rho-FoxH}) can be reduced to Meijer $G$ functions \citep{2002A&A...383..384M, 2011A&A...525A.136B}. Contrary to the Fox $H$ function, the Meijer $G$ function is available in, for example, Maple and Mathematica, so the expression above can easily be evaluated to arbitrary precision.

A very useful application of the Fox $H$ expression (\ref{rho-FoxH}) is that it enables to directly investigate the asymptotic behaviour of the luminosity density. \citet{2011A&A...534A..69B} derived the asymptotic behaviour but did not discuss its implications. As already discussed in, for example, \citetalias{1991A&A...249...99C}, the S\'ersic models for $m>1$ have a power-law asymptotic behaviour at small radii. The leading term is
\begin{equation}
\nu(r) \sim \frac{b^{3m}\,\Gamma\left(\tfrac12-\tfrac{1}{2m}\right)}{4\pi^{3/2}\,m^2\,\Gamma(2m)\,\Gamma\left(1-\tfrac{1}{2m}\right)}\,\frac{L}{\Re^3}\,
u^{1/m-1}.
\end{equation}
For large values of the S\'ersic index, the model has a $r^{-1}$ central density cusp, similar to the \citet{1990ApJ...356..359H} or NFW \citep{1997ApJ...490..493N} models, and this cusp gradually softens to a logarithmic cusp for the exponential model $m=1$. For $m<1$, the values we focus on in this paper, the S\'ersic models have a finite luminosity density, given by
\begin{equation}
\nu_0 = \frac{b^{3m}\,\Gamma(1-m)}{\pi^2\,\Gamma(1+2m)}\,\frac{L}{\Re^3}.
\end{equation}
Very interesting is the leading term in the asymptotic expansion, 
\begin{equation}
\label{nuasy}
\frac{\nu(r)}{\nu_0} \sim 
\begin{cases}
\;1+\dfrac{\Gamma(1-3m)}{2\,\Gamma(1-m)}\,u^2
&0<m<\tfrac13, \\[1.2em]
\;1+\dfrac{3\ln\tfrac{u}{2}+\gamma+\tfrac32}{2\,\Gamma(\tfrac32)}\,u^2
&m=\tfrac13, \\[1.2em]
\;1+\dfrac{\sqrt{\pi}\,\Gamma\left(\tfrac12-\tfrac{1}{2m}\right)}{2m\,\Gamma(1-m)\,\Gamma\left(1-\tfrac{1}{2m}\right)}\,u^{1/m-1} 
&\tfrac13<m<\frac12, \\[1.4em]
\;1-u^2
&m=\tfrac12, \\[0.8em]
\;1- 
\dfrac{\sqrt{\pi}\,\Gamma\left(\tfrac32-\tfrac{1}{2m}\right)}{\Gamma(2-m)\,\Gamma\left(1-\tfrac{1}{2m}\right)}\,u^{1/m-1}
&\tfrac12<m<1. 
\end{cases}
\end{equation}
This expansion shows that, at small radii, the luminosity density of the S\'ersic model actually increases with increasing radius for all values $m<\tfrac12$, as first noted by \citet{2001MNRAS.321..269T}. For $m<\tfrac13$ this increase is parabolic, and between $m=\tfrac13$ and $m=\tfrac12$ the slope gradually softens towards linear. This increase is visible in the right panel of Fig.~{\ref{SmallSersic-I-nu.fig}}, where the luminosity density of the S\'ersic model is shown for different values of $m$. 

An explicit expression for the luminosity density corresponding to the limit $m\rightarrow0$ can be determined by applying the deprojection recipe (\ref{deprojection}) to the surface brightness profile (\ref{Sigma0}). From Equation~(\ref{deprojection-gen}) \citep[see also][]{Bracewell1999}, one finds
\begin{equation}
\nu(r) = 
\begin{cases}
\;\dfrac{L}{2\pi^2 \Re^3} \left(2-\dfrac{r^2}{\Re^2}\right)^{-1/2}
&r<\!\sqrt{2}\,\Re. \\[0.5em] \;0&r>\!\sqrt{2}\,\Re. \end{cases}
\label{nu-Sersic0}
\end{equation}
This formula shows that the limiting S\'ersic model corresponding to $m=0$ is a spherical model with finite extent, and with a density profile that increases monotonically from a finite central value to an infinite density at the outer edge. We hence have a finite ball in which the outer layer has an infinite density, but the total luminosity is finite. The asymptotic behaviour at the centre is 
\begin{equation}
\nu(r) \sim \nu_0\left(1+\tfrac12\,u^2\right),
\end{equation}
in agreement with the equation~(\ref{nuasy}). This limiting case is interesting from a mathematical point of view, but has no direct astrophysical relevance. 

\subsection{Luminosity profile}
 
The luminosity profile, that is the total luminosity emitted from a sphere with radius $r$, is given by 
\begin{equation}
L(r) = 4\pi\int_0^r \nu(r')\,r'^2\,{\text{d}}r'.
\end{equation}
For the S\'ersic model, \citet{2011A&A...534A..69B} showed that it can also be written in terms of the Fox $H$ function,
\begin{equation}
L(r) = \frac{2L}{\sqrt{\pi}\, \Gamma(2m)}\,u^2\,
H^{2,1}_{2,3} \left[ \left.\begin{matrix} (0,1), (0,1) \\ (0,2m), (\tfrac12,1), (-1,1) \end{matrix}\,\right| u^2 \right].
\label{M-FoxH}
\end{equation}
From this general expression, the asymptotic behaviour at small $r$ can easily be derived. As all S\'ersic models with $m<1$ are characterised by a finite central luminosity density, it is no surprise that $L(r)$ increases as $r^3$ at small radii,
\begin{equation}
L(r) \sim \frac{2\,\Gamma(1-m)}{3\pi\,m\,\Gamma(2m)}\,L\,u^3
\quad 0<m<1.
\label{Lasy}
\end{equation}
In the limit $m\rightarrow0$, the luminosity profile becomes
\begin{equation}
L(r) = 
\begin{cases}
\;\dfrac{L}{\pi}\left[\arcsin\left(\dfrac{r}{\sqrt{2}\Re}\right)
-\dfrac{r \sqrt{2 \Re^2-r^2}}{\Re^2}\right]
&r<\!\sqrt{2}\,\Re, \\[0.5em] 
\;L&r>\!\sqrt{2}\,\Re. \end{cases}
\label{Lr-Sersic0}
\end{equation}
The asymptotic behaviour of this luminosity profile at small radii is 
\begin{equation}
L(r) \sim \frac{\sqrt2\,L}{3\pi}\left(\frac{r}{\Re}\right)^3 = \frac{4}{3\pi}\,L\,u^3,
\end{equation}
in agreement with the limit $m\rightarrow0$ of expression~(\ref{Lasy}).

\section{Dynamical properties}
\label{Dynamical.sec}

Assuming a constant mass-to-light ratio, the luminosity density $\nu(r)$ can be converted to a mass density $\rho(r)$. From this mass density profile, the gravitational potential can be calculated and the dynamical structure can be investigated. This has been done in quite some detail for the S\'ersic models with $m\geqslant2$ in \citetalias{1991A&A...249...99C}, and here we extend this analysis to the smaller values of the S\'ersic index.

\begin{figure*}
\centering
\includegraphics[width=\textwidth]{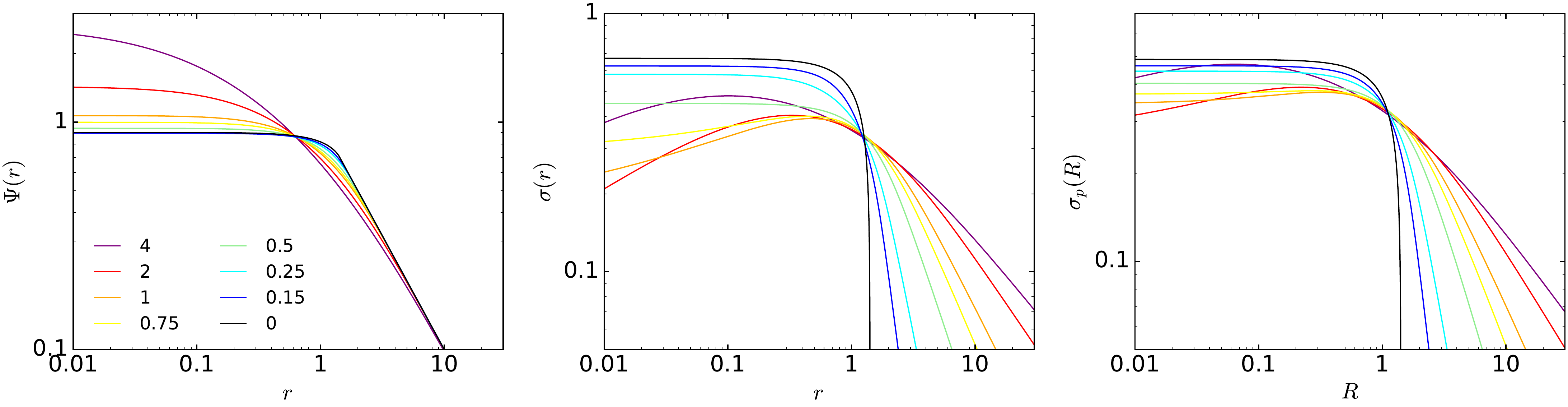}
\caption{Left: Gravitational potential $\Psi(r)$ for the same set of S\'ersic models as in Fig.~{\ref{SmallSersic-I-nu.fig}}. Middle: Velocity dispersion profile $\sigma(r)$, corresponding to an isotropic velocity distribution. S\'ersic models with $m<1$ have a finite central velocity dispersion, models with $m\geqslant1$ have a central velocity dispersion that goes to zero. Right: Observed or projected velocity dispersion profile $\sigma_{\text{p}}(R)$, again assuming an isotropic velocity distribution. All models have a finite projected velocity dispersion.}  
\label{SmallSersic-sigma.fig}
\end{figure*}

\subsection{Gravitational potential}

Given a spherical density profile $\rho(r)$, the (positive) gravitational potential $\Psi(r)$ can be determined as
\begin{equation}
\Psi(r) = 4\pi G
\left[ \frac{1}{r}\int_0^r \rho(r')\,r'^2\,{\text{d}}r' + \int_r^\infty \rho(r')\,r'\,{\text{d}}r' \right].
\end{equation}
For the S\'ersic model, the potential can be written in terms of the Fox $H$ function \citep{2011A&A...534A..69B}, 
\begin{equation}
\Psi(r) = \frac{b^m}{\sqrt{\pi}\,\Gamma(2m)}\,\frac{GM}{\Re}\,u\,
H^{2,1}_{2,3} \left[ \left.\begin{matrix} (0,1), (0,1) \\ (0,2m), (-\tfrac12,1), (-1,1) \end{matrix}\,\right| u^2 \right].
\label{Psi-FoxH}
\end{equation}
Obviously, the potential for all S\'ersic models has a Keplerian decline at large radii, and a finite central value (\citetalias{1991A&A...249...99C}; \citealt{2011A&A...534A..69B}),
\begin{equation}
\Psi_0 = \frac{2\,b^m\,\Gamma(1+m)}{\pi\,m\,\Gamma(2m)}\,\frac{GM}{\Re}.
\label{Psi0}
\end{equation}
At large $m$, the central potential is an increasing function of $m$. This is not the case for the entire range of S\'ersic parameters, however: the minimum value of $\Psi_0$ is obtained for $m\approx0.148$ (see Table~{\ref{bm.tab}}). In the limit $m\rightarrow0$, the gravitational potential can be calculated exactly as
\begin{equation}
\Psi(r) =
\begin{cases}
\;\dfrac{GM}{r}
\left[\dfrac{2}{\pi} \arcsin\left(\dfrac{r}{\sqrt{2}\Re}\right)
+
\dfrac{r \sqrt{2\Re^2-r^2}}{\pi\,\Re^2}\right]
&r<\!\sqrt{2}\,\Re ,
\\[2em] \;\dfrac{GM}{r}
 &r>\!\sqrt{2}\,\Re. \end{cases}
\end{equation}
At $r=0$, this potential reduces to
\begin{equation}
\Psi_0 = \frac{2\!\sqrt2}{\pi}\,\frac{GM}{\Re},
\end{equation}
in agreement with the limit $m\rightarrow0$ of expression~(\ref{Psi0}).

\subsection{Velocity dispersions}

Assuming an isotropic velocity distribution, the intrinsic velocity dispersion $\sigma(r)$ can be calculated using the solution of the Jeans equation,
\begin{equation}
\sigma^2(r)=\frac{1}{\nu(r)}\int_r^\infty \frac{\nu(r')\,M(r')\,{\text{d}}r'}{r'^2}.
\label{sigma2}
\end{equation}
An analytical expression for $\sigma(r)$ cannot be given, not even in terms of the Fox $H$ function, so we have to integrate this expression numerically. The central panel of Fig.~{\ref{SmallSersic-sigma.fig}} shows the velocity dispersion profile for different values of the S\'ersic parameter. At large radii, the dispersion profile is relatively shallow for large $m$, and becomes gradually steeper as $m$ decreases. As already noted in \citetalias{1991A&A...249...99C}, the velocity dispersion profiles for S\'ersic models with $m>1$ are all characterised by a central depression, with the depression being more severe for larger values of $m$. This behaviour is expected for models in which the density behaves as a power law at small radii \citep[e.g., see Appendix~C of][]{2002A&A...386..149B}. Figure~{\ref{SmallSersic-sigma.fig}} also suggests that the S\'ersic models with $m<1$ do not show this central depression, but have a finite central velocity dispersion, in line with the fact that S\'ersic models with $m<1$ have a finite central density. In Appendix~{\ref{CentralDispersion.sec}} we derive an explicit expression for $\sigma_0$ in terms of the Fox $H$ function,
\begin{multline}
\sigma_0^2 
= 
\frac{2m\,b^m}{\Gamma(2m)\,\Gamma(1-m)}\,\frac{GM}{\Re}\\ \times
H^{2,3}_{4,4} \left[ \left.\begin{matrix} (1,2m), (\tfrac12,1),(0,1),(0,1) \\ (0,2m), (\tfrac12,1), (-1,1), (1,1) \end{matrix}\,\right| 1 \right].
\label{sigma02}
\end{multline}
This expression is indeed finite for all $m<1$, and the numerical values are tabulated in Table~{\ref{bm.tab}}. Clearly, $\sigma_0$ increases for decreasing $m$. In the limit $m\rightarrow0$, we can explicitly calculate the entire velocity dispersion profile by substituting expressions~(\ref{nu-Sersic0}) and (\ref{Lr-Sersic0}) into the expression (\ref{sigma2}), 
\begin{equation}
\sigma^2(r) =
\begin{cases}
\;\dfrac{GM}{r}
\,\dfrac{2\Re^2-r^2}{\pi\,\Re^2}
\arcsin\left(\dfrac{r}{\sqrt{2}\Re}\right)
&r<\!\sqrt{2}\,\Re,
\\[0.5em] \;0
 &r>\!\sqrt{2}\,\Re. \end{cases}
\end{equation}
Setting $r=0$ in this expression we find a very simple expression for the central dispersion,
\begin{equation}
\sigma^2_0 = \frac{\sqrt2}{\pi}\,\frac{GM}{\Re}.
\end{equation}
This expression is in agreement with expression~(\ref{sigma02}) for small values of $m$ (see Table~{\ref{bm.tab}}).

From an observational point of view, the projected velocity dispersion or line-of-sight velocity dispersion $\sigma_{\text{p}}(R)$ is an important property. It can be obtained by projecting the velocity dispersion along the line of sight,
\begin{equation}
\sigma_{\text{p}}^2(R) = \frac{2}{I(R)} \int_R^\infty \frac{\nu(r)\,\sigma^2(r)\,{\text{d}}r}{\sqrt{r^2-R^2}},
\end{equation}
or equivalently
\begin{equation}
\sigma_{\text{p}}^2(R) = \frac{2}{I(R)}\int_R^\infty \frac{\nu(r)\,M(r) \sqrt{r^2-R^2}\,{\text{d}}r}{r^2}.
\label{sigmap2}
\end{equation}
Not surprisingly, an analytical solution cannot be obtained for general values of $m$. The right panel of Fig.~{\ref{SmallSersic-sigma.fig}} shows the line-of-sight velocity dispersion profiles for different values of $m$. At large radii, the line-of-sight dispersion profiles have more or less the same behaviour as the intrinsic dispersion profiles. The main difference is seen at small radii: where the intrinsic dispersion has a central depression for all $m>1$, the projected dispersion profiles converge to a finite value for all $m$. This value can be calculated explicitly (Appendix~{\ref{CentralDispersion.sec}}),
\begin{multline}
\sigma_{\text{p},0}^2 = 
\frac{4m\,b^m}{\pi\,\Gamma(2m)}\,
\frac{GM}{\Re}\\ \times
H^{2,3}_{4,4} \left[ \left.\begin{matrix} (1-m,2m), (0,1),(0,1),(0,1) \\ (0,2m), (\tfrac12,1), (-1,1), (\tfrac12,1) \end{matrix}\,\right| 1 \right].
\label{sigmap0}
\end{multline}
The central line-of-sight velocity dispersion increases with decreasing $m$, and reaches a finite value in the limit $m\rightarrow0$,
\begin{equation}
\sigma_{\text{p},0}^2 = \frac{2\!\sqrt{2}}{\pi^2}\,(2{\mathcal{C}}-1)\, \frac{GM}{\Re},
\label{sigmap00}
\end{equation}
where ${\mathcal{C}} \simeq 0.91595...$ is Catalan's constant. This expression can be found by combining the expressions~(\ref{nu-Sersic0}), (\ref{Lr-Sersic0}) and (\ref{sigmap2}); a closed expression for the line-of-sight dispersion profile for arbitrary $R$ cannot be found. Table~{\ref{bm.tab}} shows that the expression (\ref{sigmap0}) converges to the value (\ref{sigmap00}) in the limit $m\rightarrow0$.

\subsection{Distribution function}

\begin{figure*}
\centering
\includegraphics[width=\textwidth]{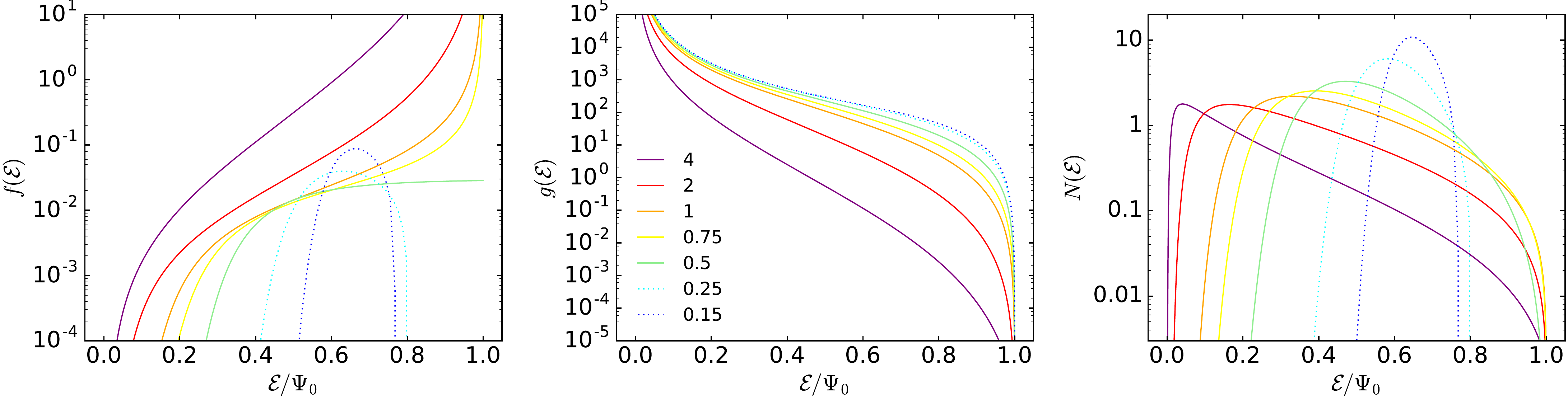}
\caption{Left: Isotropic distribution function $f({\cal{E}})$ for the same set of S\'ersic models as in Fig.~{\ref{SmallSersic-I-nu.fig}}. Only for $m\geqslant\tfrac12$ the distribution function is positive over the entire range of binding energies, and hence physical. For models with $m<\tfrac12$, the distribution function is negative for binding energies larger than some critical value, and hence these models cannot be supported by an isotropic velocity distributions. For these models, the (positive part of) the distribution function is shown as a dotted line. Middle: Density of states function $g({\cal{E}})$. Right: Differential energy distribution $N({\cal{E}})$.}
\label{SmallSersic-df.fig}
\end{figure*}

The velocity dispersion profiles derived in the previous section are only meaningful if the isotropic distribution function is positive. It is well-known that there is a unique isotropic distribution function for each spherical potential-density pair, and that this distribution function depends only on the (positive) binding energy per unit mass ${\cal{E}}$ \citep[e.g.,][]{2008gady.book.....B}. This distribution function $f({\cal{E}})$ can be calculated using Eddington's formula,
\begin{equation}
f(\calE) = \frac{1}{\sqrt{8}\,\pi^2}\int_0^{\calE}
\frac{\txd^2\rho(\Psi)}{\txd\Psi^2}\,\frac{\txd\Psi}{\sqrt{\calE-\Psi}},
\label{eddington}
\end{equation}
where $\rho(\Psi)$ is the mass density written as a function of the gravitational potential. The distribution function obtained in this way is not automatically guaranteed to be positive for $0<{\cal{E}}\leqslant\Psi_0$, which is a necessary condition for the model to be physically meaningful. 

For the S\'ersic models characterised by the potential-density pair (\ref{rho-FoxH})--(\ref{Psi-FoxH}), the Eddington formula~(\ref{eddington}) cannot be evaluated in an analytical way, so we have to use numerical means. For this purpose, expression~(\ref{eddington}) can be rewritten as
\begin{equation}
f(\calE) = \frac{1}{\sqrt{8}\,\pi^2}\int_{r(\calE)}^\infty
\frac{\Delta(r)\,\txd r}{\sqrt{\calE-\Psi(r)}},
\label{eddington2}
\end{equation}
where $r(\calE)$ is the maximum radius reachable by a star with binding energy $\calE$, defined implicitly by the equation $\Psi(r(\calE)) = \calE$, and 
\begin{equation}
\Delta(r) = 
\frac{r^2}{GM(r)}
\left[ \frac{\txd^2\rho(r)}{\txd r^2} + \frac{\txd\rho(r)}{\txd r}\left(\frac{2}{r}-\frac{4\pi\,\rho(r)\,r^2}{M(r)}\right)\right].
\end{equation}
\citet{1982MNRAS.200..951B} used this approach to numerically calculate the distribution function of the de Vaucouleurs model ($m=4$), and showed that it is everywhere positive. In \citetalias{1991A&A...249...99C}, this analysis was extended, and it was shown that this is the case for all S\'ersic models, at least with $m>2$.

We extended this analysis even more, and calculate the distribution function for S\'ersic models with arbitrary S\'ersic index using expression~(\ref{eddington2}), in particular focussing on the smaller values of $m$. We plot $f({\cal{E}})$ in the left panel of Fig.~{\ref{SmallSersic-df.fig}}. A first important result is that $f({\cal{E}})$ is positive for all $m\geqslant\frac12$. For all $m>\tfrac12$ the distribution function is a strongly increasing function of the binding energy, with $f(0)=0$, and a divergence for ${\cal{E}}\rightarrow\Psi_0$. For the gaussian model, corresponding to $m=\tfrac12$, the distribution function is also an increasing function, but the distribution function now converges towards a finite value at ${\cal{E}}=\Psi_0$. 

The situation is quite different for $m<\frac12$: for these models, the isotropic distribution function is {\em{not}} positive everywhere. In particular, the distribution function as obtained from equation~(\ref{eddington2}) is positive at small binding energies, that is at large radii, but becomes negative for all binding energies larger than some critical value ${\cal{E}}_{\text{crit}}$. For increasingly smaller $m$, ${\cal{E}}_{\text{crit}}$ becomes increasingly smaller, or equivalently, the spatial region that corresponds to a negative distribution function becomes gradually larger. We have also plotted the distribution function, or at least the part of the distribution function that is positive, of two S\'ersic models with $m<\tfrac12$ on the left panel of Fig.~{\ref{SmallSersic-df.fig}}.  

The most important conclusion that should be drawn from this observation is that all S\'ersic models with $m\geqslant\tfrac12$ can be supported by an isotropic velocity dispersion tensor, since the corresponding distribution function $f({\cal{E}})$ is positive over the entire range of binding energies. Moreover, as the distribution functions of these $m\geqslant\tfrac12$ models are monotonic, they are linearly stable with respect to both radial and non-radial perturbations, by virtue of the Doremux-Feix-Baumann theorem \citep{1971PhRvL..26..725D, 2008gady.book.....B} and Antonov's theorem \citep{1962spss.book.....A}, respectively.

The models with $m<\tfrac12$, on the other hand, cannot be supported by an isotropic velocity dispersion tensor. This behaviour is not unexpected. In fact, our findings are a direct illustration of the results obtained by \citet{1992MNRAS.255..561C}, who showed that a radially decreasing density profile is a necessary condition for the positivity of the distribution function for isotropic spherical systems.

\subsection{Differential energy distribution}

While the distribution function does contain the full kinematical information of a dynamical system, it is important to realise that $f({\cal{E}})$ does not represent the number of stars per unit binding energy. The quantity that describes this useful characteristic is the differential energy distribution $N({\cal{E}})$. For isotropic spherically symmetric systems, $N({\cal{E}})$ can be obtained as the product of the distribution function and the so-called density of states function $g({\cal{E}})$, which represents the phase-space volume per unit binding energy \citep{2008gady.book.....B}. It is calculated as 
\begin{equation}
g({\cal{E}}) = 16\!\sqrt2\pi^2\int_{\cal{E}}^\infty
\left|\,r^2\,\frac{\txd r}{\txd\Psi}\right| \sqrt{\Psi-{\cal{E}}}\,\txd\Psi.
\end{equation}
The central panel of Fig.~{\ref{SmallSersic-df.fig}} shows the density of states function for the S\'ersic models, and the right panel shows the corresponding differential energy distributions. For the de Vaucouleurs model, corresponding to $m=4$, the differential energy distribution is well approximated by a Boltzmann distribution, over a fairly large range in binding energy. This resemblance to a Boltzmann distribution was already noted by \citet{1982MNRAS.200..951B}, and in \citetalias{1991A&A...249...99C} it was argued that, of the entire family of S\'ersic models, the one with $m=4$ was most Boltzmann-like. The right panel of Fig.~{\ref{SmallSersic-df.fig}} shows that, indeed, the Boltzmannian nature of the differential energy distribution decreases for decreasing $m$, and the peak of the $N({\cal{E}})$ distribution gradually moves towards larger ${\cal{E}}/\Psi_0$ values. As soon as $m<\tfrac12$, the differential energy distribution becomes negative for large binding energies. In the limit $m\rightarrow0$ the differential energy distribution, or at least the function $N({\cal{E}})$ calculated in the way described above, is characterised by an infinite peak at 
\begin{equation}
\frac{{\cal{E}}_{\text{max}}}{\Psi_0}
=
\frac{\Psi(\sqrt2 \Re)}{\Psi_0}
=
\frac{\pi}{4},
\end{equation}
and negative values for all ${\cal{E}} > {\cal{E}}_{\text{max}}$.

We note that this does not necessarily imply that the S\'ersic models with $m<\tfrac12$ are unphysical: it still possible to self-consistently generate these density distributions with anisotropic distribution functions that favour tangential over radial orbits.  In principle, one can always build dynamical models that only contain purely circular orbits for arbitrary density profiles (leaving aside stability issues). Whether self-consistent and stable dynamical models with a gradual preference for tangential orbits can be constructed for S\'ersic models with $m<\tfrac12$ is currently unclear. Such models have been constructed for different spherical potential-density pairs \citep[e.g.,][]{1987MNRAS.224...13D, 2002A&A...393..485B, 2007A&A...471..419B, 2013MNRAS.436.2014N, 2014MNRAS.440.2636L}.

\section{Summary}
\label{Summary.sec}

The goal of this paper was to investigate in detail the main photometric, intrinsic, and dynamic properties of the S\'ersic model, with a particular focus on the subset of models with S\'ersic index $m<1$. This small $m$ regime was not discussed in most previous studies on the S\'ersic model (e.g., \citetalias{1991A&A...249...99C}; \citetalias{1997A&A...321..724C}; \citealt{2002A&A...383..384M, 2005PASA...22..118G}). S\'ersic models with small values of $m$ are very relevant, though, in particular for low-luminosity and dwarf elliptical galaxies. Indeed, the S\'ersic index correlates with luminosity \citep{1993MNRAS.265.1013C, 2003AJ....125.2936G}, and values down to $m\approx0.4$ have been found \citep[e.g.,][]{1994MNRAS.268L..11Y, 1998A&A...333...17B, 2003ApJ...582..689M}. 

We show that there is no polynomial fitting formula for the S\'ersic scaling parameter $b$ with an absolute accuracy better than $10^{-3}$ over the interval $0\leqslant m\leqslant1$. Instead we present a different fitting formula, based on a polynomial approximation for $b^m$, that has a mean absolute accuracy of $1.1\times10^{-4}$ over this interval. We also derive the formal asymptotic expansion of $b(m)$ for $m\to0$ that can be used to compute all the analytical limiting formulae presented in this paper.

While the surface brightness profile of all S\'ersic models is finite at the centre, the corresponding luminosity density shows a rich variety, depending on the value of $m$. For $m>1$, the luminosity density is characterised by a $r^{1/m-1}$ cusp, while for $m<1$ the central luminosity density is finite. For $m<\tfrac12$, the luminosity density profile increases with increasing radius at small radii, resulting in a central luminosity density depression. All S\'ersic models have a finite potential well, and a finite central projected velocity dispersion (when assuming an anisotropic velocity distribution). The intrinsic velocity dispersion drops to zero for models with $m\geqslant1$, but assumes a finite value for $m<1$. All of these properties can be calculated analytically.

All S\'ersic models with $m\geqslant\tfrac12$ can be supported by an isotropic velocity dispersion tensor, and these isotropic models are stable to both radial and non-radial perturbations. For decreasing value of $m$, the differential energy distribution deviates more and more from a Boltzmannian distribution. These results directly extend the conclusions of \citetalias{1991A&A...249...99C}. However, we find that S\'ersic models with $m<\tfrac12$ cannot be supported by an isotropic velocity distribution, which is a logical consequence of their particular density profile.

The S\'ersic model corresponding to the limit $m\rightarrow0$ is characterised by a finite extent and a uniform surface brightness distribution. In 3D, this translates to a ball in which the density increases from the centre to an outer, infinite-density skin. Most of the intrinsic and dynamical properties can be expressed in terms of elementary functions for this extreme model. 

Finally, we note that we have made use of the analytical expressions based on the Fox $H$ function, derived by \citet{2011A&A...534A..69B}. Our analysis again demonstrates the power of the underestimated Fox $H$ function as a tool for analytical work.

\section*{Acknowledgements}

We thank Francesco D'Eugenio for interesting discussions and a careful reading of this manuscript. We thank the referee for suggestions that clarified the content and presentation of this work.

\bibliographystyle{aa}
\bibliography{SmallSersic}

\appendix

\section{Numerical values}

In Table~{\ref{bm.tab}} we present numerical values for a number of important properties of the S\'ersic models, as a function of $m$. All quantities are given in dimensionless units, or equivalently, for a model with $\Re = L = M = 1$.

\begin{table*}
\caption{Numerical values for a number of important properties of the S\'ersic models, as a function of the S\'ersic index $m$. The different columns correspond to the scaling parameter $b$, the ratio $b^m = \Re/h$, the central surface brightness $I_0$, the central luminosity density $\nu_0$, the central value of the gravitational potential $\Psi_0$, the central velocity dispersion $\sigma_0$, and the central projected velocity dispersion $\sigma_{\text{p},0}$. All quantities are given in dimensionless units, or equivalently, for a model with $\Re = L = M = 1$.}
\label{bm.tab}
\centering
\begin{tabular}{cccccccc}
\hline \hline \\
$m$ & $b$ & $b^m$ & $I_0$ & $\nu_0$ & $\Psi_0$ & $\sigma_0$ & $\sigma_{\text{p},0}$ \\ \\ \hline \\
$0.0$ & $0.0$ & 	             $0.7071067812$ & $0.1591549431$   & $0.03582244802$  & $0.9003163162$ & $0.6709382670$ & $0.4882770668$ \\
$0.1$ & $0.02074633919$ & $0.6787254708$ & $0.1597040458$   & $0.03687117924$  & $0.8954110224$ & $0.6467405417$ & $0.4745932365$ \\
$0.2$ & $0.1450781416$   & $0.6797048938$ & $0.1657441274$   & $0.04174912260$  & $0.8955714657$ & $0.6052798406$ & $0.4549379359$ \\
$0.3$ & $0.3157020170$   & $0.7075924771$ & $0.1783670066$   & $0.05214840909$  & $0.9049229140$ & $0.5576761459$ & $0.4365182562$ \\
$0.4$ & $0.5013512264$   & $0.7586768505$ & $0.1967139352$   & $0.07074452800$  & $0.9202187043$ & $0.5054510861$ & $0.4198025544$ \\
$0.5$ & $0.6931471806$   & $0.8325546112$ & $0.2206356002$   & $0.1036366539$    & $0.9394372787$ & $0.4490044158$ & $0.4046143026$ \\
$0.6$ & $0.8879362135$   & $0.9311701833$ & $0.2504981367$   & $0.1646935035$    & $0.9614737847$ & $0.3883701725$ & $0.3907268693$ \\
$0.7$ & $1.084371511$     & $1.058338681$   & $0.2870242952$   & $0.2892627286$    & $0.9857010672$ & $0.3231072572$ & $0.3779411016$ \\
$0.8$ & $1.281796137$     & $1.219706019$   & $0.3312367088$   & $0.5903862892$    & $1.011747061$   & $0.2516350723$ & $0.3660931503$ \\
$0.9$ & $1.479856427$     & $1.422976161$   & $0.3844538418$   & $1.656657118$      & $1.039383734$   & $0.1684024227$ & $0.3550500768$ \\
$1.0$ & $1.678346990$     & $1.678346990$   & $0.4483153816$   & $\infty$                   & $1.068468879$   & $0.0$                   & $0.3447039338$ \\
$1.5$ & $2.674060314$     & $4.372771642$   & $1.014407550$     & $\infty$                   & $1.233536106$   & $0.0$                   & $0.3008238298$ \\
$2.0$ & $3.672060749$     & $13.48403014$   & $2.411450297$     & $\infty$                   & $1.430700033$   & $0.0$                   & $0.2658868180$ \\
$2.5$ & $4.670908883$     & $47.15236589$   & $5.897607399$     & $\infty$                   & $1.662679605$   & $0.0$                   & $0.2367684306$ \\
$3.0$ & $5.670161189$     & $182.2998096$   & $14.69230924$     & $\infty$                   & $1.934261054$   & $0.0$                   & $0.2118423541$ \\
$3.5$ & $6.669637075$     & $766.2274547$   & $37.07959732$     & $\infty$                   & $2.251549732$   & $0.0$                   & $0.1901483907$ \\
$4.0$ & $7.669249443$     & $3459.485015$   & $94.48274697$     & $\infty$                   & $2.621876860$   & $0.0$                   & $0.1710622608$ \\
$4.5$ & $8.668951184$     & $16628.36345$   & $242.5415290$     & $\infty$                   & $3.053889794$   & $0.0$                   & $0.1541466369$ \\
$5.0$ & $9.668714615$     & $84497.47049$   & $626.2886060$     & $\infty$                   & $3.557722251$   & $0.0$                   & $0.1390769232$ \\
\\
\hline \hline \\
\end{tabular}
\end{table*}

\section{Asymptotic expansion of the scaling parameter for $m\rightarrow0$}
\label{bmzero.sec}

\begin{figure}
\includegraphics[width=0.9\columnwidth]{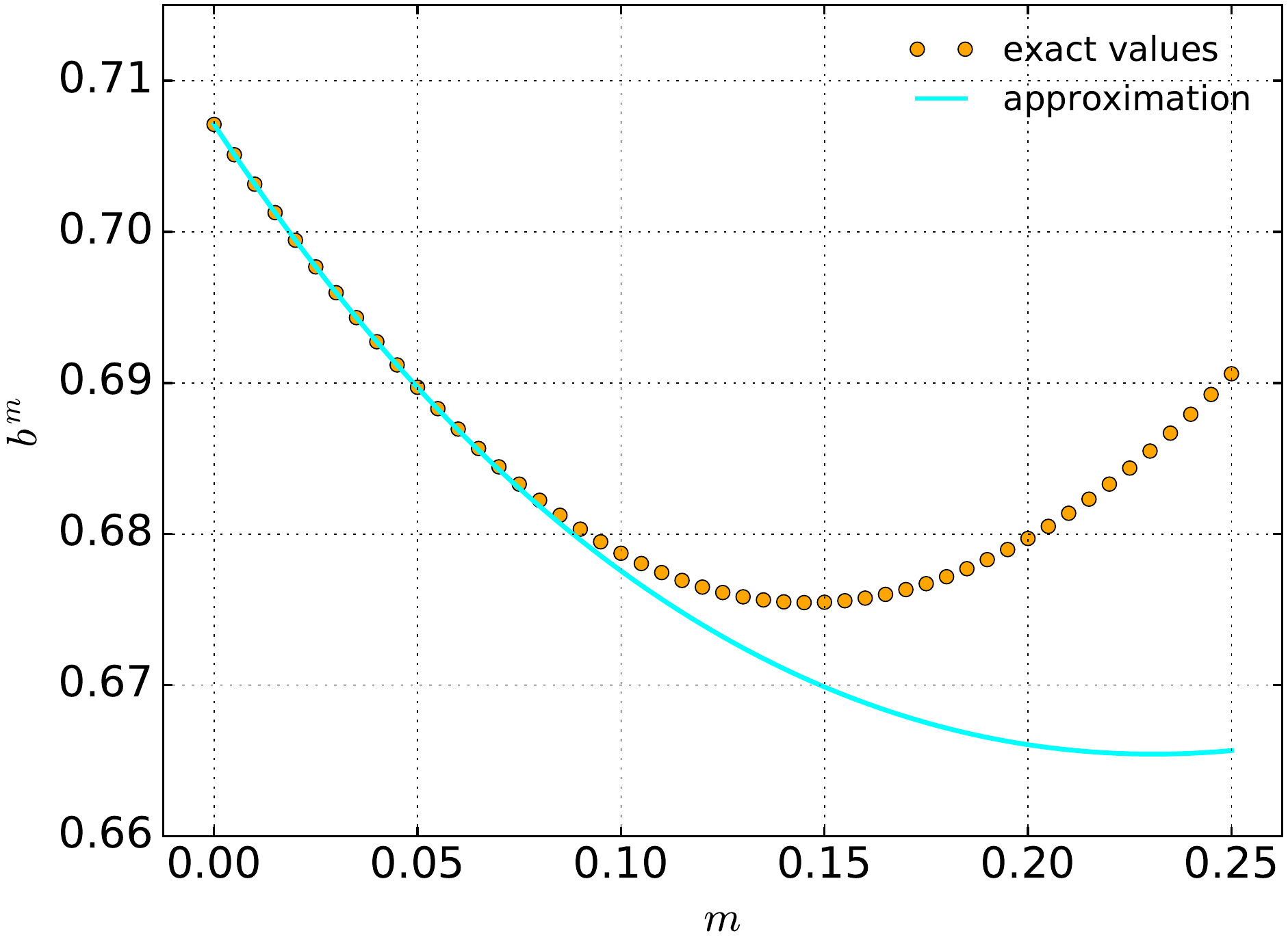}
\caption{The quantity $b^m$ for very small values of $m$. The yellow dots correspond to the exact value determined by numerical evaluation of equation~(\ref{solvebm}), the cyan line is the approximation (\ref{approx}).}  
\label{ScalingParameterAsymptotic.fig}
\end{figure}

The scaling parameter $b$ is defined through equation~(\ref{solvebm}). In order to find an asymptotic form for $b(m)$ for $m\rightarrow0$, we take the Laurent expansion of the right-hand expression,
\begin{equation}
\frac{\Gamma(2m)}{2} = \frac{1}{4m} - \frac{\gamma}{2} + \frac12\left(\frac{\pi^2}{6}+\gamma^2\right)m + {\mathcal{O}}(m^2)
\label{Xright}
\end{equation}
with $\gamma\simeq0.577216...$ the Euler-Mascheroni constant. The left-hand side of (\ref{solvebm}) can be written as
\begin{align}
\gamma(2m,b) = \int_0^b t^{2m-1}\,e^{-t}\,{\text{d}}t = \frac{b^{2m}}{2m} + \sum_{j=1}^\infty \frac{(-1)^j\,b^{2m+j}}{(2m+j)\,j!}
\label{Xleft}
\end{align}
We search for a solution for $b^{2m}$ of the form
\begin{equation}
b^{2m} = X_0 + X_1\,m + X_2\,m^2 + \cdots
\end{equation}
Inserting this expansion into expression (\ref{Xleft}) and equating the result to (\ref{Xright}), we can determine each of the coefficients by using order balance. After some calculation, one finds 
\begin{equation}
b^{2m} = \frac12 - \gamma\,m + \left(\frac{\pi^2}{6}+\gamma^2\right)m^2 + \cdots
\label{b2mseries}
\end{equation}
Repeating the treatment at all orders, it follows that the leading term of the asymptotic expansion of $b(m)$ for $m\to 0$ can be obtained in closed form as
\begin{equation}
b^m \sim \sqrt{\,m\,\Gamma(2m)}.
\label{approx}
\end{equation}
This is also shown in Fig.~{\ref{ScalingParameterAsymptotic.fig}}, where we compare the numerically determined values of $b^m$ and the approximation (\ref{approx}).

\section{Exact expressions for the central dispersions}
\label{CentralDispersion.sec}

Assuming an isotropic velocity dispersion tensor, the intrinsic velocity dispersion profile $\sigma(r)$ can be derived using the solution of the Jeans equation,
\begin{equation}
\sigma^2(r) = \frac{G}{\nu(r)}\int_r^\infty \frac{M(r')\,\nu(r')\,{\text{d}}r'}{r'^2}.
\end{equation}
In the case $m<1$, the central luminosity is finite, and we obtain for $r=0$, 
\begin{equation}
\sigma_0^2 = \frac{G}{\nu_0} \int_0^\infty \frac{M(r)\,\nu(r)\,{\text{d}}r}{r^2}.
\label{Asig0}
\end{equation}
Substituting the expressions (\ref{rho-FoxH}) and (\ref{M-FoxH}) for $\nu(r)$ and $M(r)$, this becomes
\begin{multline}
\sigma_0^2 = 
\frac{2m\,b^m}{\Gamma(2m)\,\Gamma(1-m)}\,
\frac{GM}{\Re}
\\
\qquad\quad\times
\int_0^\infty 
H^{2,1}_{2,3} \left[ \left.\begin{matrix} (0,1), (0,1) \\ (0,2m), (\tfrac12,1), (-1,1) \end{matrix}\,\right| x \right]
\\
\times
H^{2,0}_{1,2} \left[ \left.\begin{matrix} (0,1) \\ (0,2m), (\tfrac12,1) \end{matrix} \,\right| x \right]
x^{-1}\,dx.
\end{multline}
The integral in this expression is a particular case of a Mellin transform of the product of two Fox $H$ functions, which can be evaluated again as a Fox $H$ function \citep[][\S2.3]{2009hfta.book.....M}, 
\begin{multline}
\sigma_0^2 
= 
\frac{2m\,b^m}{\Gamma(2m)\,\Gamma(1-m)}\,\frac{GM}{\Re}\,
\\
\times
H^{2,3}_{4,4} \left[ \left.\begin{matrix} (1,2m), (\tfrac12,1),(0,1),(0,1) \\ (0,2m), (\tfrac12,1), (-1,1), (1,1) \end{matrix}\,\right| 1 \right].
\end{multline}
In the special case $m=\tfrac12$, corresponding to a Gaussian model, this Fox $H$ function reduces to a Meijer $G$ function, which can be expressed in terms of elementary functions,
\begin{equation}
\sigma_0^2 
= 
\sqrt{\frac{b}{\pi}}\,\frac{GM}{\Re}\,
G^{1,2}_{2,2} \left[ \left.\begin{matrix} \tfrac12, 0 \\ \tfrac12, -1 \end{matrix}\,\right| 1 \right]
= \frac{(4-\pi)\sqrt{b}}{2\sqrt{\pi}}\,
\frac{GM}{\Re}.
\end{equation}

The projected velocity dispersion profile can be found through the formula
\begin{equation}
\sigma_{\text{p}}^2(R) = \frac{2}{I(R)}\int_R^\infty \frac{M(r)\,\nu(r)\sqrt{r^2-R^2}\,{\text{d}}r}{r^2}.
\end{equation}
For $R=0$, this expression reduces to 
\begin{equation}
\sigma_{\text{p},0}^2 = \frac{2}{I_0} \int_0^\infty \frac{M(r)\,\nu(r)\,{\text{d}}r}{r}.
\end{equation}
This expression is very similar to expression (\ref{Asig0}), and can be evaluated in a similar way. The resulting expression is
\begin{multline}
\sigma_{\text{p},0}^2 = 
\frac{4m\,b^m}{\pi\,\Gamma(2m)}\,
\frac{GM}{\Re}
\\
\times
H^{2,3}_{4,4} \left[ \left.\begin{matrix} (1-m,2m), (0,1),(0,1),(0,1) \\ (0,2m), (\tfrac12,1), (-1,1), (\tfrac12,1) \end{matrix}\,\right| 1 \right].
\label{Asigp0}
\end{multline}
For the Gaussian model $m=\tfrac12$, this expression can be simplified to a Meijer $G$ function and expressed in terms of elementary functions,
\begin{multline}
\sigma_{\text{p},0}^2
= 
\frac{2\!\sqrt{b}}{\pi}\,\frac{GM}{\Re}\,
G^{1,2}_{2,2} \left[ \left.\begin{matrix} 0, 0 \\ \tfrac12, -1 \end{matrix}\,\right| 1 \right]
\\=
\frac{[2\ln(1+\sqrt2)-\sqrt{2}]\!\sqrt{b}}{\sqrt{\pi}}\,\frac{GM}{\Re}.
\end{multline}
Also for other integer and half-integer values of $m$, the Fox $H$ function in expression (\ref{Asigp0}) can be reformed to a Meijer $G$ function, now by application of Gauss' multiplication theorem. One finds after some calculation
\begin{multline}
\sigma_{\text{p},0}^2
= 
\frac{4(2m)^mb^m}{(2\pi)^{2m}\,\Gamma(2m)}\,\frac{GM}{\Re}
\\ \times
G^{2m,2m+1}_{2m+1,2m+1} \left[ \left.\begin{matrix} 
-\tfrac{m-1}{2m},-\tfrac{m-1}{2m},\ldots,\tfrac{m-1}{2m},0,0 \\[0.3em]
\tfrac{1}{2m},\tfrac{2}{2m},\ldots,\tfrac{2m-1}{2m},\tfrac12,-1
\end{matrix}\,\right| 1 \right].
\end{multline}
Setting $m=1$ in this expression, we find a surprisingly simple result for the central projected velocity dispersion of the exponential model,
\begin{equation}
\sigma_{\text{p},0}^2
= 
\frac{2b}{\pi^2}\,\frac{GM}{\Re}\,
G^{2,3}_{3,3} \left[ \left.\begin{matrix} 0, 0, 0 \\ \tfrac12, \tfrac12, -1 \end{matrix}\,\right| 1 \right]
=
\frac{(\pi-3)\,b}{2}\,\frac{GM}{\Re}.
\end{equation}

\end{document}